\documentclass[10pt]{article}
\usepackage[margin=1in]{geometry}
\usepackage[font=small]{caption}
\usepackage{amsmath,amssymb,amsthm,mathtools}
\usepackage{stmaryrd}
\usepackage{booktabs,graphicx,xcolor}
\usepackage[round]{natbib}
\usepackage[hidelinks]{hyperref}
\usepackage{microtype}

\theoremstyle{definition}

\newcommand{\state}{\mathbf{w}}            
\newcommand{\dens}{\rho}
\newcommand{\pres}{p}

\newcommand{\hmesh}{h}

\newcommand{\dt}{\Delta t}
\newcommand{\cfl}{\mathrm{CFL}}

\newcommand{\envK}{K}

\newcommand{\admset}{\mathcal{G}}          

\title{Guarantees by Construction for Learned Finite Volume Schemes\\
on Steady Supersonic Flow}
\author{Denis Gueyffier\\
Direction Scientifique G\'en\'erale, ONERA\\
Institut Polytechnique de Paris, Palaiseau, France}
\date{}

\begin{document}
\maketitle

\begin{abstract}
A second order finite volume scheme rests on two local quantities: a gradient
reconstructed in each cell, and a limiter which scales it down where the
reconstruction would overshoot. Both are set by fixed formulas, and on coarse
unstructured meshes a small network can supply better values. But a network is
free to output anything, and the usual safeguard is a penalty in the training
loss, which discourages inadmissible states without preventing them.

We replace the penalty by a hard constraint. The network still sets both
quantities, and every value it can produce lies inside bounds that are safe.
Its stencil weights cannot cancel a
neighbour, so the least squares system stays invertible. Its limiter is capped by
a value computed from the local flow, below which no reconstruction creates a new
extremum or drives density or pressure negative. The flux, the wall treatment and the time step are not learned
and carry their own guarantees. Admissibility therefore holds for every value of
the weights rather than as an outcome of training, and no negative density or
pressure occurred in any computation reported here.

Because the scheme is safe whatever the network does, we could ask what the
network contributes. We test it on supersonic channel flow over an obstacle,
including the forward facing step of Woodward and Colella. Learning lowers the
error by 38\% on an unseen geometry and 29\% on an unseen obstacle topology,
measured against the same scheme with the network switched off, which itself
outperforms a classical smooth limiter on three cases of four. The method aims at the accuracy of a
fine mesh for the cost of a cheaper coarse one, and we measured how far in this
direction it goes. Refining the mesh once improves the error
fourfold and multiplies the run time by eight. Learning secures half of this
improvement for a sixth of this time, three times as much accuracy per unit of
cost.

All of this comes from one of the two quantities the network sets. The gradient reconstruction
reproduces the full effect on its own, and the limiter accounts for about a tenth
as much. This also explains why the gain fades beyond the Mach numbers the weights were
trained on.
\end{abstract}

\section{Introduction}
\label{sec:intro}

A second order finite volume scheme requires two local quantities in every cell,
at every step: a gradient estimated from the neighbouring cells, and a limiter
that scales it down. On a triangulation the first is harder to obtain than it
looks. The three neighbours sit at arbitrary angles, some of them
possibly across a shock, and the least squares fit that produces the gradient
weighs them all alike. The limiter is the classical dilemma of second order
schemes. Use the full gradient and the reconstruction may overshoot near a
discontinuity, driving the density or the pressure below zero and stopping the
computation. Use too little and the scheme falls back to first order accuracy
where nothing was wrong.

A line of work asks whether a network can set the gradient and the limiter better
than a fixed formula does, cell by cell, and it can
\citep{deromemont2025,deromemont2026}.
The difficulty is that a network is free to output anything. That freedom is what
makes it useful, and it is also what can produce the overshoot the limiter exists
to prevent. The usual safeguard adds terms to the training loss that
penalize entropy production and spurious oscillations. Such penalties act where
the training data went, and they discourage rather than prevent. The same work states
the resulting gap plainly. A finite volume method that is efficient, accurate on
coarse meshes and above all robust, while maintaining physical properties such as
the entropy inequality, remains an open problem.

\paragraph{What we do.} We close this gap by changing the nature of the constraint. Instead of penalizing inadmissible outputs, we
put the network's outputs inside bounds that no inadmissible value can reach, and
we let it choose freely there. The
gradient weights are bounded away from cancellation, so the least squares system
stays invertible whatever the network produces. The limiter is capped by a value
computed from the local flow, below which every choice yields a reconstruction
that creates no new extremum and keeps density and pressure positive. Around this
learned core, the rest of the solver carries its own guarantees: an entropy
stable flux, a wall condition that makes the boundary entropy flux vanish by
symmetry, and a time step given by an explicit formula that keeps the updated
cell averages positive. The network never
produces a flux, an update, or a time step.

What this yields is a learned scheme with hard guarantees rather than a learned
component with a safety net. Admissibility holds for every value of the weights,
at initialization and after every gradient step. It is a property of the
construction, not an outcome of training. Across every computation reported in
this paper, on all geometries, all seeds and all parameter values we swept, no
negative density or pressure was ever produced.

\paragraph{What we found.} Because the scheme is safe whatever the network does,
we could ask what the network actually contributes, and get a clean answer. It
turns out to be sharper than we expected, and directly actionable. Of the two quantities the network sets, only one carries
the gain. Freezing the learned
limiter and keeping only the learned gradient weights reproduces the full effect
of the network, to within a tenth of a percent on the cases where it helps.
Freezing the gradient weights and keeping only the learned limiter moves the
error by three percent instead of thirty. \emph{What learning adds to this solver
is a better gradient, not a better limiter.}

The method aims at the accuracy of a fine mesh for the cost of a cheaper coarse
one, and we measured how far in this direction it goes. On our reference case,
refining the mesh once improves the error fourfold and multiplies the run time by
eight. Learning secures half of this improvement for a sixth of this time, three
times as much accuracy per unit of cost.

That single fact organizes the rest of the paper. It explains why the relaxation
parameter of our envelope, which we found to be badly tuned, changes the accuracy
without changing what the network learns. It explains why restricting the network
to a band near the walls helps on familiar flows and hurts on unfamiliar ones.
And it says something about the construction itself: the guarantee constrains the
limiter while the gain comes from the gradient, so the two never compete.

\paragraph{Where we test it.} On supersonic channel flow over an obstacle,
including the forward facing step of \citet{woodward1984}, a standard benchmark
whose expansion corner is a known difficulty. Walls are present throughout
training, as they are in any real computation, and the evaluation holds out an
unseen geometry, two unseen Mach numbers, and an unseen obstacle topology.
Section~\ref{sec:method} gives the construction in full, and
Section~\ref{sec:protocol} the cases and the rules we fixed before measuring.
Section~\ref{sec:results} reports what the method delivers, and
Section~\ref{sec:negatifs} what the steady regime changes about training.

\section{What has been learned, and what has been guaranteed}
\label{sec:related}

\paragraph{Learned finite volumes for compressible flow.} The line this work continues replaces the fixed
slope limiter of a second order finite volume scheme by learned decisions. It covers
Cartesian grids with varying boundary conditions \citep{deromemont2026} and
unstructured triangulations with a learned gradient reweighting
\citep{deromemont2025}. Admissibility is pursued there through
regularizers added to the loss, which penalize entropy production and total
variation. That work states the open problem plainly: a finite volume method that
is efficient, accurate on coarse meshes, interpretable and above all robust while
maintaining physical properties such as the entropy inequality remains
unresolved. The present paper addresses exactly that gap, by enforcing
admissibility in the way the reconstruction is built rather than through the
loss.

\paragraph{Learned solvers and their evaluation habit.} The founding line runs
from data-driven discretizations of partial differential equations
\citep{barsinai2019} through learned advection schemes \citep{zhuang2021} to
solvers trained through the simulator \citep{kochkov2021}. Those works
established the coarse mesh promise and the practice of training through unrolled
trajectories. They share an evaluation habit worth naming: the domain is
periodic, and generalization is demonstrated along the forcing and the Reynolds
number, never along the boundary. \citet{kochkov2021} also popularized reporting
that accounts for computing cost, which we turn here into thresholds fixed in
advance.

\paragraph{Unrolled training.} Training on unrolled trajectories reduces the
mismatch between the states seen during training and those produced at inference,
because the observed samples converge towards the learned attractor
\citep{um2020,kohl2022}. The same works report that long horizons destabilize the
gradient, and that a curriculum on the horizon is needed
\citep{list2024}. A variant without temporal gradients unrolls
the trajectory but propagates no gradient across steps. It comes close to the
fully differentiable one while storing no intermediate state, so its memory does
not grow with the horizon. All of this literature
addresses transient or chaotic dynamics, which matters for the interpretation of
our own negative result in Section~\ref{sec:negatifs}.

\paragraph{Deciding where to limit.} A limiter that acts everywhere pays for
robustness in regions that never needed it, and a large literature therefore aims
to restrict it. The distinction usually sought is not between shocks and
smooth flow but between monotone and non-monotone behaviour. A smooth extremum
can be recognized by the local monotonicity of the derivative, whereas a spurious
oscillation cannot, and a marker applied to the second derivative separates the
two. Monotonicity preserving schemes \citep{sureshhuynh1997} and the
multi-dimensional limiting process with its smooth extremum detector
\citep{park2012} build on this idea, the latter on unstructured grids. Recent
work pushes it further, classifying the field into smooth, weakly non-smooth and
discontinuous regions and reconstructing each differently \citep{mpl2024}.

We considered adding such a detector to switch the relaxation of
Section~\ref{sec:garanties} on and off locally, and Section~\ref{sec:negatifs}
reports why we did not. On the flows studied here the relaxation degrades the
smooth regions more than the shocked ones, so a detector that enabled it away
from shocks would apply it precisely where it costs most. Strict clipping has a
compensating virtue that is worth naming, since it keeps what the limiter
constrains aligned with what a detector would flag.

\paragraph{Two ways to keep a scheme admissible.} Outside machine learning, the
question of how to prevent a high order reconstruction from producing
inadmissible states has two established answers, and our construction belongs
squarely to one of them. The first acts \emph{a priori}: a limiter shrinks the
reconstruction before it is used, so that no forbidden state is ever formed. That
is the line of \citet{barthjespersen1989}, of the accuracy preserving variants
\citep{michalak2009}, and of the positivity scaling of \citet{zhangshu2010}. The
second acts \emph{a posteriori}: the update is computed optimistically, then
checked against admissibility and stability criteria, and recomputed with a lower
polynomial degree wherever the check fails. That is the MOOD paradigm
\citep{clain2011,diot2012}, which explicitly includes density and pressure
positivity among its detection criteria.

The two philosophies trade the same quantities in opposite directions. Detecting
afterwards keeps full accuracy wherever nothing goes wrong, at the price of
recomputing troubled cells and of carrying a fallback scheme. Building safely
beforehand pays a fixed cost everywhere and never recomputes, but constrains the
reconstruction even where the constraint was unnecessary. Our reason for choosing
the first is specific to learning. A network is only safe by construction if the
set it outputs into is safe, which requires the constraint to be known
\emph{before} the network acts. An a posteriori guarantee would hold for the
solver but not for the network, and the distinction matters precisely under the
distribution shift that learned schemes are meant to survive. A learned scheme
built on a MOOD backbone is a natural counterpart to this work, and we have not
attempted it.

\paragraph{Learning on steady compressible flow.} A large body of work predicts
steady transonic and supersonic fields directly from geometry and flow
conditions, without any time integration \citep{thuerey2020}. These are
surrogates rather than solvers, so the question of unrolling does not arise for
them. Closer to the present setting, \citet{feng2024} augment a discontinuous
Galerkin solver with a learned model that supplies an initial field, and report a
reduction of the number of iterations needed to reach the steady state. Their
learning acts once, before the computation, and their gain is one of convergence
speed. Ours acts at every cell and every step, and its gain is one of accuracy at
a given resolution. The two objectives are distinct, and we pursue the second.

\section{A learned scheme that cannot leave the admissible set}
\label{sec:method}

The scheme is a classical finite volume solver in which a network is allowed to
act, and the construction is defined by what the network may not do. It reweights
the gradient stencil and picks one number per cell inside an interval that is
proved admissible. It never touches the flux, the update, or the time step.
The flux, the update and the step bound are therefore fixed code, identical
whether the network is trained, untrained, or absent, which is why admissibility
holds for any weights.

\subsection{Governing equations}
We solve the two-dimensional compressible Euler equations for an ideal gas,
\begin{equation}
\partial_t \state + \nabla \cdot \mathbf{F}(\state) = 0,
\qquad
\state = (\dens,\ \dens v_x,\ \dens v_y,\ E)^{\top},
\label{eq:euler}
\end{equation}
with $E = \pres/(\gamma-1) + \tfrac{1}{2}\dens|\mathbf{v}|^2$ and $\gamma = 1.4$
throughout. The admissible set is
$\admset = \{\state : \dens > 0,\ \pres(\state) > 0\}$, and preserving it is what
the construction below enforces. The specific entropy is
$s = \ln \pres - \gamma \ln \dens$, and a scheme is entropy stable when the
discrete total mathematical entropy does not increase in the absence of boundary
fluxes \citep{tadmor1987}.

\subsection{The solver the network sits in}
Cell averages are advanced on unstructured triangulations by a second order
finite volume method. Each cell $K_i$ stores one average $\bar q_i$
per variable. Second order accuracy requires a linear profile inside the cell,
\begin{equation}
q_i(\mathbf{x}) = \bar q_i + \phi_i\, \mathbf{g}_i \cdot (\mathbf{x} - \mathbf{x}_i),
\label{eq:reconstruction}
\end{equation}
where $\mathbf{x}_i$ is the centroid, $\mathbf{g}_i$ a gradient estimated from the
neighbours, and $\phi_i \in [0,1]$ the limiter. Setting $\phi_i = 0$ recovers a
first order scheme, robust and diffusive; setting $\phi_i = 1$ uses the full
gradient, accurate and liable to overshoot. The gradient is obtained by weighted
least squares over the three face neighbours $j$,
\begin{equation}
\mathbf{g}_i = M_i^{-1} \sum_{j} w_{ij}\, (\bar q_j - \bar q_i)\, \mathbf{d}_{ij},
\qquad
M_i = \sum_j w_{ij}\, \mathbf{d}_{ij} \mathbf{d}_{ij}^{\top},
\label{eq:lsq}
\end{equation}
with $\mathbf{d}_{ij}$ joining the two centroids and $w_{ij}$ a geometric weight.
On a triangulation the three neighbours sit at arbitrary angles, so this estimate
is far less well conditioned than a one-dimensional slope, which is one reason
classical limiters are conservative on unstructured meshes. Boundary faces carry one of three types.
A slip wall uses the mirror state with respect to the face normal. A supersonic
inflow imposes the upstream state, since every characteristic enters the domain.
A supersonic outflow extrapolates from the interior, since none does. The last
two follow \citet{deromemont2025}.

\paragraph{Time integration.} Time advances explicitly with a step common to the whole domain. It follows a first order positivity bound evaluated on the current state, $\dt = \cfl \min_i |K_i| / (P_i s_i)$. Here $P_i$ is the cell perimeter, $s_i$ the local wave speed, and $\cfl = 0.3$. This is an explicit sufficient bound
with a safety factor, not the largest admissible step, since the admissible set
is open and a largest step need not exist. This formula is the classical
sufficient bound for a first order scheme with Rusanov dissipation. Our chain
uses an entropy stable flux and a limited reconstruction, so the bound is applied
outside the setting in which it is proved, and Section~\ref{sec:garanties}
reports the margin we measure instead. Cell averages are updated by a two increment
scheme in the spirit of \citet{berthon2005,berthon2006}, who show that a second order scheme written as a convex average of first order updates inherits the invariant domain of those updates under a suitable step restriction. Let $R(\state)$ denote the limited residual of
Section~\ref{sec:garanties}. The update is then
\begin{equation}
\state^{(1)} = \state^{n} + \dt\, R(\state^{n}),
\qquad
\state^{n+1} = \tfrac{1}{2}\big(\state^{n} + \state^{(1)} + \dt\, R(\state^{(1)})\big).
\label{eq:update}
\end{equation}
At a constant step this is Heun's method, and what distinguishes it here is that
the intermediate state $\state^{(1)}$ passes through the same admissibility
construction as $\state^{n}$. The scheme therefore never evaluates the residual
at an inadmissible state, which a plain two stage method does not guarantee. The
explicit choice with a step common to the domain is inherited from the reference
implementation; Section~\ref{sec:limites} explains why a steady problem would be
better served by a local one.

\subsection{What the network does}
The learned component does not replace the solver, it informs it. The
architecture is the geometric operator head of \citet{deromemont2025}, and its
inputs are invariant by construction. A branch reads the twelve differences of
primitive variables to the three neighbours, which makes it invariant under
translation of the state. A trunk reads three angular descriptors of the stencil
geometry, which makes it invariant under rotation of the mesh. The two are
combined by an elementwise product and a linear head, with dimensions
$12 \to 28 \to 28$ for the branch and $3 \to 28$ for the trunk, GELU activations
throughout, and 1404 parameters in total. Its size is dictated by its use: it is
evaluated at every cell and every step, so inference must be cheap for the scheme
to remain competitive. It produces four numbers per cell, and each enters the
scheme at one precise place.

Three of them, written $\alpha_{i1}, \alpha_{i2}, \alpha_{i3}$, reweight the
least squares stencil of~\eqref{eq:lsq}, whose weights become
$w_{ij}(1+\alpha_{ij})$. The network can thus lean on one neighbour rather than
another, which matters when a discontinuity crosses the stencil and one neighbour
carries information from the wrong side of it. Each $\alpha$ is bounded in
$[-\tfrac12, \tfrac12]$ by $\alpha = \tfrac12 \tanh(\cdot)$, so a weight can be
halved or increased by half but never cancelled or reversed, and $M_i$ stays
invertible. That bound also controls conditioning, which matters on triangular
stencils. We swept the eight corners of the reachable cube on
the canonical step mesh. The worst attainable condition number of $M_i$ is $14.2$
against $10.1$ without reweighting, a degradation of at most $2.5$ per cell and
$2.0$ at the median. The network cannot make the gradient fit appreciably worse conditioned
than the geometry already makes it. On boundary cells the three are forced to zero, so that the
reweighting cannot interfere with the treatment of the wall.

The fourth output sets the limiter position $\phi_i$
of~\eqref{eq:reconstruction} through a sigmoid $\lambda_i \in (0,1)$, and the next
section explains why it is a position inside an interval rather than a value in
$[0,1]$.

That fourth output deserves a remark. In its original form the sigmoid starts at one half. An untrained network therefore begins by using half of the gradient it is allowed whereas the backbone uses all of it. The regularizer then pulls the bias back towards that same half. We shift the sigmoid argument by
a constant, which places both the starting point and the resting point of the
regularizer at the backbone. The parametrization, the number of
parameters and the class of representable functions are unchanged. Measured on
one seed at equal budget, the shift lowers the error by 18.5\%.

\subsection{Training through the solver}
The solver is written in JAX, and so is the network. The composition of the two
is differentiable, and the training gradient is obtained by reverse mode through
the discretization itself rather than through a surrogate. That
matters here more than it would for a smooth scheme. The chain is full of
non-smooth operations: the minimum over faces in the limiter, the clipping
against the envelope, the scaling towards the cell average. Automatic
differentiation handles them in the almost everywhere sense, and we checked that
the result is the derivative one wants. On a training pair, the reverse mode
gradient of the loss with respect to the 1404 parameters agrees with a finite
difference of the same derivative to within $5 \times 10^{-7}$ relative.

This is what makes the training loss a function of the solver rather than of a
proxy for it: the network is fitted against the discretization it will run
inside, with the limiter and the positivity scaling in the loop.

\subsection{Admissibility by construction}
\label{sec:garanties}
An unconstrained learned scheme can leave $\admset$, and does so under
distribution shift. The construction removes that failure mode by design rather
than by penalty, in three nested steps.

\paragraph{The interval lemma.} The Barth--Jespersen limiter
\citep{barthjespersen1989} answers a simple question. How much of the gradient
can be used before a face value leaves the range of the neighbouring averages? Let
$q^{\min}_i$ and $q^{\max}_i$ be the smallest and largest averages over the cell
and its neighbours, and let $\delta_{if} = \mathbf{g}_i \cdot (\mathbf{x}_f -
\mathbf{x}_i)$ be the unlimited increment at face $f$. The largest admissible
factor at that face is
\begin{equation}
\psi_{if} =
\begin{cases}
(q^{\max}_i - \bar q_i)/\delta_{if} & \text{if } \delta_{if} > 0,\\
(q^{\min}_i - \bar q_i)/\delta_{if} & \text{if } \delta_{if} < 0,\\
1 & \text{otherwise,}
\end{cases}
\qquad
\phi_{\mathrm{BJ},i} = \min\big(1, \min_f \psi_{if}\big).
\label{eq:bj}
\end{equation}
Since $\bar q_i$ lies in $[q^{\min}_i, q^{\max}_i]$, every $\psi_{if}$ is
nonnegative, so $\phi_{\mathrm{BJ},i} \in [0,1]$. The key point is that this is a
ceiling and not a value. For any $\phi \in [0, \phi_{\mathrm{BJ},i}]$, every
reconstructed face value lies between the local minimum and maximum of the
neighbouring averages. The admissible set of
slope factors is therefore the whole interval $[0, \phi_{\mathrm{BJ}}]$. Any
learned output mapped into it yields a reconstruction that creates no new
extremum. The learned limiter thus parameterizes a position inside a provably
safe envelope, the unstructured analogue of a Sweby region, rather than an
unconstrained correction.

\paragraph{The relaxed envelope.} A strict Barth--Jespersen envelope has a known
weakness. Consider a smooth extremum, a maximum of the flow
that is not a discontinuity. The cell average is already the largest of its
neighbourhood, so $\psi_{if}$ collapses and the scheme drops to first order where
nothing is wrong.
The classical remedy widens the bounds by an amount that vanishes under
refinement \citep{michalak2009}, and we use
\begin{equation}
q^{\min}_i - \envK \hmesh_i^2 \ \le\ q_i(\mathbf{x}_f) \ \le\
q^{\max}_i + \envK \hmesh_i^2,
\qquad \hmesh_i = \sqrt{2 A_i},
\label{eq:relaxed}
\end{equation}
with $A_i$ the cell area. The tolerated overshoot is $O(\hmesh^2)$, so it
vanishes in the refinement limit and the maximum principle is recovered
asymptotically, while second order accuracy is restored at smooth extrema on a
given mesh. The parameter $\envK$ controls how much overshoot is tolerated, and
$\envK = 0$ recovers the strict envelope.

We swept $\envK$ over $0$, $0.5$, $1$ and $2$ on the canonical step, computing a
separate fine reference at each value so that every scheme is compared to its own
refinement limit. Two things come out of it, and they point in opposite
directions.

No negativity event occurs at any value, including $\envK = 2$. The guarantee is
therefore independent of this parameter, which is what the construction predicts.
Relaxation acts on the envelope that forbids new extrema, whereas positivity is
enforced by the separate scaling described below, and the sweep confirms that the
two are decoupled.

Accuracy, on the other hand, is best at $\envK = 0$. The error per cell rises
monotonically with the parameter: $0.64$ at $\envK = 0$, $2.03$ at $\envK = 0.5$,
$2.37$ at $\envK = 1$. On a flow dominated by shocks the relaxation buys little,
since smooth extrema are rare there, and it pays for the overshoot it tolerates.
We therefore set $\envK = 0$ throughout this paper, and keep the term in the
formulation because it costs nothing and matters on smoother problems.

Accuracy is another matter, and the same sweep is unflattering. The error per
cell rises monotonically with $\envK$, from $0.64$ at $\envK = 0$ to $2.03$ at
our value of $0.5$ and $2.37$ at $1$. On a flow dominated by shocks, the
relaxation buys little, since smooth extrema are rare there, and it pays for the
overshoot it tolerates. We report the sweep rather than retune, since $\envK$
enters every measurement in this paper. Section~\ref{sec:limites} names it as the
first thing we would change next. The relaxation vanishes as $\hmesh \to 0$, so the envelope is
preserved in the limit while second order accuracy is restored away from
discontinuities. Inside it, $\phi_{\mathrm{BJ}}$ is the exact minimum of the per
face constraints, at training as at evaluation. Those constraints are nonnegative
by construction, so $\phi_{\mathrm{BJ}} \in [0,1]$ without any further clamping,
and the interval lemma applies during training exactly as it does at inference.
The learned limiter is then $\phi = \lambda\, \phi_{\mathrm{BJ}}$, with
$\lambda \in (0,1)$ the sigmoid output. The network chooses a position strictly
inside a provably safe interval, and cannot leave it at initialization or after
any gradient step.

\paragraph{Positivity of the updated averages.} Reconstruction admissibility does
not by itself guarantee that the updated averages stay positive. The scaling limiter of \citet{zhangshu2010} pulls reconstructed states towards the cell average until density and pressure are positive. A third scaling shrinks them further, so that the intermediate state of the two increment scheme is admissible as well. Each of the three steps only shrinks, so the properties obtained by the
previous ones are preserved by convexity. What remains is a bound on the time step. For a
first order scheme with Rusanov dissipation, the classical argument gives such a
bound as an explicit formula, and that is the one we use. Our flux is not of that
form, but it is close enough to it that the argument can be recovered under one
explicit condition.

Write the entropy stable flux as a local Lax--Friedrichs flux plus a remainder,
\begin{equation}
\mathbf{F}^{\mathrm{ES}}
= \underbrace{\tfrac{1}{2}\big(\mathbf{F}(\state_L)+\mathbf{F}(\state_R)\big)
- \tfrac{1}{2} s_{\max} \llbracket \state \rrbracket}_{\text{local Lax--Friedrichs}}
+\ \mathbf{D},
\qquad
\mathbf{D} = \mathbf{F}^{\mathrm{EC}} - \tfrac{1}{2}\big(\mathbf{F}(\state_L)+\mathbf{F}(\state_R)\big).
\label{eq:decomposition}
\end{equation}
The classical positivity argument applies to the first group. It survives the
addition of $\mathbf{D}$ as long as the net dissipation stays positive, which
holds if $\mathbf{D}$ is dominated by the dissipation it sits next to. Measuring
that domination in the form that governs the entropy balance, with
$\llbracket \mathbf{w} \rrbracket$ the jump of the entropy variables, the
condition reads
\begin{equation}
\big| \llbracket \mathbf{w} \rrbracket \cdot \mathbf{D} \big|
\ \le\ \theta \cdot \tfrac{1}{2} s_{\max}\,
\llbracket \mathbf{w} \rrbracket \cdot \llbracket \state \rrbracket,
\qquad \theta < 1 .
\label{eq:theta}
\end{equation}
The right hand side is positive by convexity of the entropy, and both sides are
bilinear in the jump, so the ratio stays bounded as the jump vanishes. Under
\eqref{eq:theta} the classical bound holds with $s_{\max}$ replaced by
$s_{\max}/(1-\theta)$, that is, with the time step divided by $1-\theta$.

The condition is weaker than the conclusion, since it bounds a flux discrepancy
while the conclusion concerns positivity of cell averages. It is also
checkable.
Measured over a full 15233 step run on the forward facing step, $\theta$ peaks at
$0.242$ and its per sample maxima have median $0.198$. The bound is therefore
divided by at most $1.32$, so the Courant number of $0.3$ we use corresponds to
an effective $0.396$, still below one. Our step is covered, with a factor of four
of margin on $\theta$ itself. What is general here is the statement; the
verification is one case and one trajectory. Along every trajectory reported here,
the largest step that keeps all averages admissible is at least fourteen times
the one actually taken, at every probed state. We report that margin as a margin.

\paragraph{Scope.} The guarantee is a property of the chain as specified,
interior flux and step bound included, and we probed that scope rather than
assuming it. Replacing the entropy stable flux by a Rusanov flux and changing
nothing else produces one inadmissible cell after about two thousand steps on a
wall bounded case. The failing cell carries a wall face, sits in a stagnant near
vacuum pocket, and has its limiter clamped to zero, so neither the learned heads
nor the relaxed envelope is active there. We read this as a property of the flux
and of the boundary treatment near vacuum rather than of the learned
construction, and we state the guarantee for the chain we ship.

\subsection{Why admissible states are not enough}
The interior flux is an entropy stable pair in the sense of Tadmor
\citep{tadmor1987}. It combines two ingredients. The first is the
entropy conservative flux of \citet{ismailroe2009}, built from logarithmic means
of the $\mathbf{z}$-vector variables. The second is Rusanov dissipation applied
to the jump of the conservative state,
\begin{equation}
\mathbf{F}^{\mathrm{ES}} = \mathbf{F}^{\mathrm{EC}}
- \tfrac{1}{2}\, s_{\max}\, \llbracket \state \rrbracket,
\label{eq:esflux}
\end{equation}
with $s_{\max}$ the maximal wave speed at the face. The mapping from entropy
variables to conservative ones is the gradient of a convex potential, so the jump
product is nonnegative and the dissipation term can only destroy entropy. The
pair therefore satisfies a discrete cell entropy inequality. At wall faces the
mirror state makes the entropy flux vanish by symmetry, which extends the
inequality to solid boundaries.

Two scope notes belong here. The inequality is a first order statement, and the
limited reconstruction with the learned limiter is not covered by the proof. We
therefore pair it with an executable contract, checked on every build. It
requires that a two hundred step rollout on a wall case leave the total entropy
nonincreasing to $10^{-10}$. And the entropy stable core costs 7.7\% per step
over a plain Rusanov flux, which is part of the 16\% reported below. Admissibility of
the states and stability of the scheme are distinct properties, and the
distinction is not academic. A scheme can keep every cell
admissible to the last step and still blow up, with the kinetic energy growing
while the minimum pressure collapses. The entropy stable core closes that gap.
The measured convergence order of the full chain is 2.33.

\section{What we varied, and how we decided}
\label{sec:protocol}

\subsection{What counts as a case}
Claiming that a scheme generalizes requires saying what was held fixed and what
was varied, and on a wall bounded problem that is less obvious than it looks. Two
runs described as different geometries may also differ in their boundary types
and in their initial state, so that a difference in error cannot be attributed to
any one of the three. We therefore treat a case as a triple: a geometry, a set of
boundary condition types, and an initial condition. Every case below shares the same
boundary types, so that geometry and flow regime can be varied one at a time.

The common frame is a channel of length three and height one, with a supersonic
inflow on the left, a supersonic outflow on the right, and slip walls above,
below, and on the obstacle. The initial state is uniform at the inflow
conditions, with $\gamma = 1.4$, density $1.4$ and pressure $1$, so that the
sound speed is one and the Mach number is the inflow velocity. The coarse mesh
has about two thousand cells, obtained by explicit triangulation of a structured
grid whose lines contain the obstacle edges, followed by a perturbation of the
strictly interior nodes. The boundary is left untouched, so the domain area is
exact to machine precision.

\subsection{Twelve cases to train on, three to be judged on}
Twelve cases are used for training. Eight carry a forward facing step, with
heights $0.15$ and $0.25$, positions $0.5$ and $0.8$, and inflow Mach numbers $2$
and $3$. Four carry a bump, that is a block of finite length, which presents a
descending step as well as an ascending one. Neither the height $0.20$ nor the
position $0.60$ appears among them.

\begin{figure}[htbp]
\centering
\includegraphics[width=\textwidth]{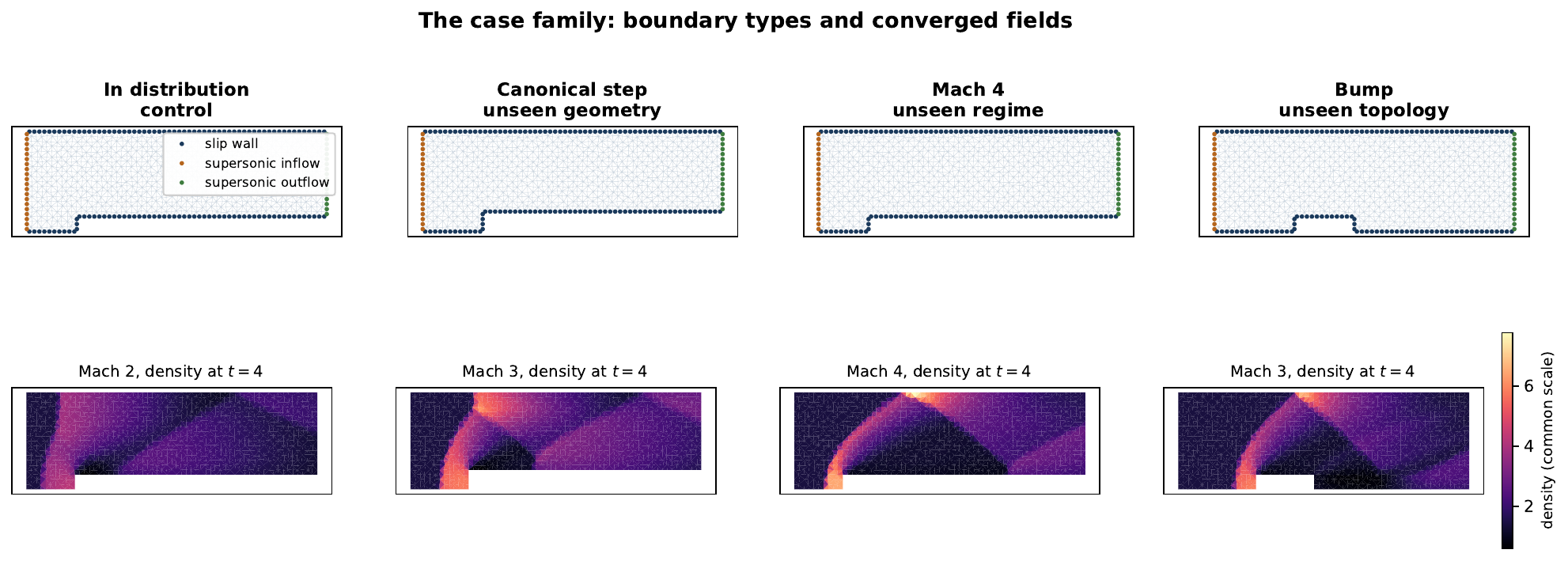}
\caption{The case family. Top: the geometry and the type of every boundary face.
Bottom: the density at $t = 4$, on a scale common to all four panels, so that the
cases can be compared with one another. The control is drawn from the training
set; the other three are held out, one per axis of generalization.}
\label{fig:famille}
\end{figure}

Four cases are held out, covering three axes of generalization, and none shares its
parameters with the training set. The geometry axis is the canonical forward
facing step of \citet{woodward1984}, at height $0.20$ and position $0.60$: both
values are unseen, and the case carries published reference contours. The regime
axis is a training geometry evaluated at Mach $4$ and at Mach $5$, neither of them seen in training. The topology axis
is a bump whose height and position are both seen, so that the only unseen
feature is the finite length. Table~\ref{tab:cas} lists the whole family, and Figure~\ref{fig:famille} shows the four evaluation cases. A fourth case, drawn from the training set, serves
as an in-distribution control. If learning brings nothing there, the training
itself is at fault rather than the axes.

\begin{table}[htbp]
\centering
\begin{tabular}{llccccc}
\hline
Case & Role & Height & Position & Length & Mach & Cells \\
\hline
Steps, 8 cases   & training & 0.15, 0.25 & 0.5, 0.8 & $\infty$ & 2, 3 & 1900--2136 \\
Bumps, 4 cases   & training & 0.15, 0.25 & 0.5, 0.8 & 0.4, 0.6 & 2, 3 & 2280--2352 \\
In distribution  & control & 0.15 & 0.5 & $\infty$ & 2 & 2100 \\
Canonical step   & unseen geometry & \textbf{0.20} & \textbf{0.60} & $\infty$ & 3 & 2016 \\
Mach 4 case      & unseen regime & 0.15 & 0.5 & $\infty$ & \textbf{4} & 2100 \\
Mach 5 case      & unseen regime & 0.15 & 0.5 & $\infty$ & \textbf{5} & 2100 \\
Bump             & unseen topology & 0.15 & 0.8 & \textbf{0.6} & 3 & 2328 \\
\hline
\end{tabular}
\caption{The case family. Every case shares the same boundary types, namely a
supersonic inflow, a supersonic outflow and slip walls, so that geometry and
regime vary one at a time. Values in bold are absent from the training set. The
control is drawn from the training set; the last three are held out, one per axis
of generalization.}
\label{tab:cas}
\end{table}

\subsection{What the error is measured against}
These flows reach a steady state, so the quantity of interest is the converged
field rather than a trajectory. That is a claim we measure rather than assume.
We take the relative rate of change
$\|\state^{n+1} - \state^{n}\| / (\dt \|\state^{n}\|)$ as an indicator. Between the start of each run and $t = 4$ it drops by a
factor of $32$ to $278$ on the evaluation cases. On eleven of the twelve training
cases it drops by a factor of $12$ to $91$. The twelfth,
a tall step close to the inflow at Mach $2$, only drops by a factor of $2.8$ and
is not steady at $t = 4$. We keep it in the training set, since its pairs are
physically correct and a spread of convergence rates is useful there, and we
report it here rather than averaging it away. Errors are measured at $t = 4$ against a reference computed on a nested hierarchy. The coarse mesh is refined once to give the fine level, and once more to give the reference level, so that each level contains the previous one. Projection from one level to the previous one is the
area weighted average, which is exact since every parent has four children of a
quarter of its area. We checked that the reference mass is conserved through both
projections to the ninth digit. The error itself is the discrete $L^1$ distance in primitive variables,
\begin{equation}
E = \frac{1}{N} \sum_{i=1}^{N} \sum_{v \in \{\dens,\, v_x,\, v_y,\, \pres\}}
\big| q_v(K_i) - q_v^{\mathrm{ref}}(K_i) \big|,
\label{eq:error}
\end{equation}
summed over the four primitive variables and over the $N$ cells of the coarse
mesh, then divided by $N$. The four variables enter unweighted and in their
physical units, which the uniform inflow state normalizes to order one. Dividing by $N$ matters when meshes of different
sizes are compared, since a summed norm grows with the number of terms. In
Section~\ref{sec:isocout} the summed error would suggest that refinement degrades
the solution, whereas per cell it improves by a factor $2.21$.

\subsection{How we decided what counts}
A learned scheme offers many ways to be accidentally right, so the rules were
fixed before looking. Each campaign was written down in advance: the arms, the
cases, the seeds, the horizon, the thresholds that would decide the outcome, and
where it applied, a clause stating what result would count against us. Verdicts
are reported literally against those thresholds, negative ones included. When
first contact with the data forced a change, the amendment is dated and kept
beside the original rather than folded into it.

Two conventions follow from this. An axis on which the sign of the effect changes
between seeds is reported as undetermined, whatever the sign of its median. And
numbers obtained under a scheme that we later replaced are either remeasured or
labelled as such.

\section{What the method delivers}
\label{sec:results}

\subsection{The scheme without a network}
\label{sec:sans-apprentissage}
Before asking what learning contributes, one has to know what it is being added
to. Set the gradient weights to zero and place the limiter at the top of its
proved interval, and the construction of Section~\ref{sec:garanties} still
defines a complete scheme, with no learned parameter anywhere. We call it the
backbone, and it is the reference against which the learned contribution is
measured.

It completes every case in this study. The forward facing step combines a supersonic
inflow, a supersonic outflow, slip walls and the expansion corner that makes the
case difficult. On it, the backbone runs to $t = 4$ in 15233 steps without a
single negative density or pressure. Two conditions underlie the entropy argument at the
boundary, and both are verified rather than assumed: at a supersonic outflow the
normal velocity must exceed the sound speed, and at a supersonic inflow it must
lie below its negative. Measured at every step and every boundary face, over more
than a million face and step pairs, neither is ever violated.

The corner deserves a word, since it is the known difficulty of this case. An
erroneous entropy layer forms along the wall downstream of it, and our
measurements reproduce it. The specific entropy excess is $0.45$ just upstream of
the corner, jumps to $0.76$ just downstream, and stays between $0.71$ and $0.78$
along the entire downstream wall. The level mixes the physical entropy rise
across the bow shock with the spurious contribution of the corner; it is the jump
of about $0.30$ that is attributable to the corner. Entropy stability does not
remove that artefact, which is consistent with what entropy stable discontinuous
Galerkin schemes report on the same geometry.

\subsection{How the backbone compares to a classical limiter}
\label{sec:venkat}
A learned scheme measured only against its own unlearned version proves little,
since that version can always be made weak. We therefore replaced our interval limiter by the
smooth limiter of \citet{venkatakrishnan1995}, changing nothing else: same
entropy stable flux, same boundaries, same step bound, same mesh, same reference.
Both arms here are purely numerical, with no learned parameter anywhere, so the
comparison is between two ways of choosing $\phi$ inside the same guaranteed
chain. The four flows below are the same ones used later
for the learned arm. For two classical schemes they carry no notion of being seen
or unseen. This compares two ways of choosing $\phi$ inside the same guaranteed
chain, not two solvers, and Table~\ref{tab:venkat} gives the outcome.

\begin{table}[htbp]
\centering
\begin{tabular}{lrrr}
\hline
Flow & Backbone & Venkatakrishnan & Difference \\
\hline
Step $h = 0.15$ at Mach 2 & \textbf{0.1545} & 0.2674 & $-42$\% \\
Step $h = 0.20$ at Mach 3 & \textbf{0.5748} & 0.7132 & $-19$\% \\
Step $h = 0.15$ at Mach 4 & \textbf{0.2880} & 0.4431 & $-35$\% \\
Bump $h = 0.15$ at Mach 3 & 0.5341 & \textbf{0.4642} & $+15$\% \\
\hline
\end{tabular}
\caption{Error per cell against the converged fine reference. Neither arm
contains a network, so these are four distinct flows rather than a distribution
and its complement, and they are named by their geometry and Mach number. The
backbone is ahead on three of them, by 19 to 42\%, and it wins those three
against a handicap. The reference is generated by a scheme using the same smooth
limiter, which favours that arm.}
\label{tab:venkat}
\end{table}

The backbone is ahead on three cases of four. That is enough to make it a fair
reference for what follows, which is the only claim we need here.

\subsection{What learning adds, and where it comes from}
\label{sec:apports}
Against that backbone, the trained network lowers the error by 38\% on an unseen
geometry and by 29\% on an unseen obstacle topology. It raises it by 11\% in
distribution and by 90\% at an unseen Mach number. The gains and the losses are
both large, and neither is a rounding effect.

The interesting question is which of the two learned decisions produces them. The
network outputs gradient weights and a limiter position, and they can be
separated: freeze one at its unlearned value and keep the other.

\begin{table}[htbp]
\centering
\begin{tabular}{lrrr}
\hline
Case & Gradient weights only & Limiter only & Both \\
\hline
In distribution & $+13.4$\% & $+7.8$\% & $+9.0$\% \\
Unseen geometry & $\mathbf{-38.4}$\% & $-3.2$\% & $-38.3$\% \\
Unseen Mach 4 & $+95.3$\% & $+3.3$\% & $+89.7$\% \\
Unseen topology & $\mathbf{-28.8}$\% & $-3.2$\% & $-28.7$\% \\
\hline
\end{tabular}
\caption{Relative change of the converged error against the backbone, with each
learned output enabled separately. All three arms use trained weights, so the
flows are named here by what they hold out from training, unlike
Table~\ref{tab:venkat}. The gradient weights alone reproduce the full effect:
$-38.4$ against $-38.3$, and $-28.8$ against $-28.7$. The limiter alone moves the
error ten times less.}
\label{tab:decomposition}
\end{table}

The answer is unambiguous. The learned gradient weights alone reproduce what the
full network does, on the cases where it helps and on the cases where it hurts.
The learned limiter contributes about a tenth as much, in the same direction.
What the network adds to this solver is a better estimate of the gradient,
obtained by leaning on the neighbours that carry relevant information and away
from those that do not.

This has a consequence for the construction. The guarantee acts on the limiter,
capping it below a value computed from the local flow. The gain comes from the
gradient weights, which are bounded but not otherwise constrained. The two therefore act on different
quantities, and making the scheme safe costs nothing in accuracy. We could
have assumed this; instead we measured it.

\subsection{The limit of the transfer}
Reading the same table the other way shows where the method stops working.
Everything travels through the gradient weights, and those weights are learned
from flows at Mach 2 and 3.
Applied at Mach 4 they raise the error by 90\%, and at Mach 5 by 81\%: same sign,
same magnitude, so this is a trend and not an isolated point. Leaning on
neighbours the way a Mach 3 flow rewards is the wrong thing to do at Mach 5.

Geometry behaves differently. An unseen step height and position, and an unseen
obstacle topology, both benefit. The stencil geometry there is the same as in
training even though the shape is not, so the learned reweighting still applies.
The boundary of the transfer, on this training set, is the flow regime rather
than the shape.

\subsection{How the gain depends on mesh size}
\label{sec:echelle}
A practitioner will ask what happens on the finer meshes of a real computation.
The network reads local differences and stencil angles, so it applies at any
resolution without retraining, and the question is decidable. We ran it at three
levels, each compared to a reference four times finer, which is the ratio used in
training, and Table~\ref{tab:echelle} collects the result.

\begin{table}[htbp]
\centering
\begin{tabular}{lrrrr}
\hline
Cells & Reference & Backbone & Learned & Gain \\
\hline
504 & 2016 & 0.9367 & 0.7380 & $-21.2$\% \\
\textbf{2016} & \textbf{8064} & 0.5748 & 0.3546 & $\mathbf{-38.3}$\% \\
8064 & 32256 & 0.4274 & 0.3743 & $-12.4$\% \\
\hline
\end{tabular}
\caption{The learned gain against the backbone at three mesh sizes, each measured
against a reference four times finer. The middle row is the resolution the
network was trained on. The gain peaks there and falls off on both sides.}
\label{tab:echelle}
\end{table}

The gain is largest at the resolution the network was trained on and smaller both
above and below it. That is a real limitation, and it says what the learned
correction is: not a relative improvement that transports across resolutions, but
a setting tuned to a mesh size. The network reads differences of primitive
variables whose magnitude scales with the cell size. A stencil four times coarser
or finer therefore presents it with inputs outside the range it was fitted on. This
is consistent with the decomposition above, since the gain travels entirely
through the gradient weights, which act on exactly those magnitudes.

The obvious remedy is to retrain at the target resolution, and we tried it. The
result points somewhere else: a network trained at 8064 cells scores $0.4660$
there, behind both the backbone and the network trained at 2016 and applied
unchanged.
One difference between the two runs is large enough to explain this on its own.
Generating a training target at 8064 cells requires a converged solution at
32256, which costs about as much as the reference of this table. The retrained
network therefore saw one case where the other saw twelve. What the comparison establishes is therefore about data
rather than resolution. Twelve cases at the wrong mesh size beat one case at the
right one. Whether retraining at scale would help given
a full training set is a question we leave open, and answering it needs twelve
converged solutions at 32256 cells.

\subsection{What the gain is worth in mesh refinement}
\label{sec:isocout}
The aim of this line of work is the accuracy of a fine mesh at the price of a
coarse one, so the gains above are better read in units of classical refinement
than as percentages. Establishing that conversion requires meshes that are
actually nested. Two meshes of the same size generated independently differ by
their interior node perturbation, and comparing them measures that difference
rather than a resolution effect. We therefore refine one coarse mesh twice, so
that each level contains the previous one, and project by area weighted averages,
which are exact since every parent has four children of a quarter of its area.

On that hierarchy the classical curve is monotone: going from 2016 to 8064 cells
costs $7.9$ times more wall-clock time and divides the error per cell by $2.21$,
a slope of $-0.38$ in logarithmic scale. The learned arm costs 16\% more per run
than the backbone. Spent on refinement instead, that budget buys a 5.5\% error
reduction, whereas learning delivers 38\% and 29\% on the two axes where it
helps. Put in the terms that matter to a practitioner. Refining the mesh once takes the
error per cell from $0.5748$ to $0.1424$, and the run from 52 to 416 seconds.
Learning secures half of that improvement for 69 seconds, a sixth of what the
fine mesh costs, so each second buys three times as much accuracy. Read the other
way, reaching the same accuracy by refinement alone would take 106 seconds
against 69, so learning is $1.5$ times cheaper at equal accuracy.

One reservation belongs with this figure. The classical slope is measured on the
forward facing step and assumed to hold on the other cases. It is an
extrapolation, defensible because the slope of a convergence curve varies little
within one family of geometries, and stated rather than hidden.

\section{What the steady regime changes}
\label{sec:negatifs}

Most of the machinery developed for learned solvers was designed for transient
and chaotic flows, where the solution explores an attractor and never settles.
Our flows do settle. Three techniques that the literature recommends turn out to
be unnecessary in that regime, and the reason is the same in each case: they
correct a difficulty that a converging flow does not have.

\paragraph{Relaxing the envelope away from shocks would not help.} The natural
refinement of our construction is to keep the strict envelope near
discontinuities and relax it elsewhere, driven by one of the detectors of
Section~\ref{sec:related}. We measured whether that has a target before building
it, and it does not.

We classified cells by the steepness of the pressure gradient on the converged
reference, taking the steepest 15\% as shocked and excluding their immediate
neighbours as a safety band. Of the error of the strict scheme, 46\% sits in the
smooth cells and 38\% in the shocked ones, the remainder falling in the excluded
band. That looks like an opportunity. But comparing the same two schemes within that group
settles it the other way. Going from $\envK = 0$ to $\envK = 0.5$ raises the
error by 331\% in the smooth cells, against 104\% in the shocked ones. The relaxation does not merely miss its target, it does most damage
where it was meant to help, so a detector enabling it away from shocks would
apply it exactly where it costs most.

Why the smooth cells behave this way is a question we can only answer partially.
They are smooth in the sense of a pressure gradient sensor. But they carry the
expansion fan and the shear layer shed by the corner, where a tolerated overshoot
is amplified downstream rather than corrected. The relevant distinction on this
flow is not shocked against smooth but monotone against non-monotone, and a
gradient based sensor does not capture it. A detector built on the monotonicity
of the derivative might; we did not pursue it, since the mechanism it would
enable is the one we measured to be harmful here.

\paragraph{The training budget is not the binding constraint.} Doubling the number of epochs from four
to eight degrades the result on all three seeds. The clearest signal is not on
the generalization axes but on the in-distribution control, where learning turns
from a gain on all three seeds into a loss on all three. A network that gets
worse in distribution as it trains longer is not short of data, and is not
failing to generalize; it is optimizing a quantity that is not the one being
measured. The training losses show the same thing directly: they reach their
minimum partway through and rise afterwards, on every seed, while the rollout
quality decays.

Tripling the amount of data, at a fixed number of epochs, also degrades, though
more gently. We had expected the two levers to act in opposite directions, since
more epochs repeat the same samples while more data diversify them. The two
levers do not act in opposite directions.
With 1404 parameters, the network reaches what it can extract quickly, and
further input of either kind moves it towards an average solution that fits
individual cases less well. Capacity, rather than data, is the candidate binding
constraint, and we leave that untested.

\paragraph{Unrolled training is not needed in this regime.} Training on unrolled
trajectories is the standard remedy for the mismatch between the states seen
during training and those produced at inference. We tried both of its usual forms. The fully
differentiable version, which backpropagates through nine steps, required a batch
eight times smaller to fit in memory and still degraded the control. The version without temporal gradients fits in memory at the nominal batch size, which allows a fair comparison. It also degrades the control, from a gain of 8.0\% to a loss of 11.9\%. It does improve the unseen geometry, where it
gives the best value of the study.

The literature recommends raising the unrolling horizon progressively rather than
imposing it at once \citep{list2024}. Such a curriculum was tried in this family
of schemes, at horizons of two, five and ten steps and on three seeds, and it did
not change the outcome in any material way.

The explanation we favour is structural. Unrolling reduces a data mismatch that is small here to begin with. Our flows converge to a fixed point, so the states the model produces and the states it was trained on approach the same limit. The
literature on unrolling addresses transient and chaotic dynamics, where the
attractor is explored rather than reached. What justifies the method elsewhere is therefore
already supplied by the steady regime itself, while its cost in gradient
stability remains. Consistently, the unrolled arm helps most on the axis furthest from the
training set and hurts on the control. We also note that
\citet{deromemont2025} state that unrolled steps are not needed for robustness in
this family of schemes; our measurements support that statement rather than
contradict it.

\paragraph{A periodic pre-training stage is unnecessary when walls are in the data.} The reference work
trains first on purely periodic data and then fine-tunes on wall bounded data,
and reports that without this staging the method tends to diverge on walls. We
reproduced the staging at equal total budget, and it degrades the control by
almost twenty points relative to direct training. The reason is visible in the
difference between the two settings: their periodic stage supplies what their
main dataset lacks, namely walls. Ours contains walls throughout, so the periodic
stage fills no gap and spends half the budget on a geometry and a set of
boundaries that are never evaluated. The result does not contradict their recipe;
it shows that the recipe is empty once the training set already contains walls.

\section{How far this goes, and where it should go next}
\label{sec:limites}

\paragraph{Scope.} The construction and its guarantee are general: they rest on
the interval lemma and on convexity, not on the flow regime or on the geometry.
What the measurements below establish is narrower, and the boundary is worth
stating. Two of the four generalization axes are undetermined across seeds, and
we do not claim them. The classical cost curve is measured on
one case and assumed to hold on the others. The experiments of
Section~\ref{sec:negatifs} use a single seed each, so they establish the
existence of an effect rather than its reproducibility. All results are obtained
on one coarse resolution, of about two thousand cells, and on one family of
geometries, all built from rectangular cuts of a channel. Nothing here speaks to
viscous flow, since the equations are inviscid, and the Reynolds number is
therefore outside the reach of this corpus.

\paragraph{The time integration is inherited, not optimal.} These flows reach a
steady state, and the natural measure for such a problem is the number of steps
needed to converge. We measured it and found no acceleration: learning improves
the accuracy of the state reached, not the speed at which it is reached. That is
partly a consequence of a design choice we did not revisit. The solver advances
explicitly under a stability condition with a step common to the whole domain,
which is the standard choice in this line of work, made there for training speed.
For a steady problem it is the wrong tool: fifteen thousand steps are needed here
where an implicit method with a local time step converges in a few hundred
\citep{feng2024}. Moving to a local time step is the natural next step, and the
reference work names an implicit variant as its own robustness perspective. One
reservation applies: the two increment scheme and the intermediate state
limitation both assume a step common to all cells, so the entropy argument would have to be examined again.

\paragraph{Where the binding constraint may lie.} Doubling the epochs and
tripling the data both degrade the result, which points away from the training
budget and towards the model. With 1404 parameters, chosen so that inference
remains cheap at every cell and every step, the network may reach its capacity
early. Testing that would mean varying the width while holding the per step cost
accountable, and it is the experiment we would run next.

\paragraph{Reproducibility.} All computations reported here run on a single core of an Intel Xeon at 2.80 GHz
with 3 GB of memory, in double precision, without GPU or parallelism. The
absolute timings depend on that, the ratios between them do not. The software is
JAX 0.11.0 with optax 0.2.8 under Python 3.12.3. The scheme, the case generator, the frozen protocol sheets and the evaluation scripts form a single tree under version control. Every campaign is recorded there with the thresholds fixed before it was run, and the verdict returned against them. Each result reported here carries the identifier of the sheet that produced
it. Amendments made after first contact with the data are dated and kept beside
the original rather than folded into it; three such amendments occur in this
paper, and each is stated where it applies. Reference solutions and trained weights are
archived with the code, since neither can be reproduced in a reasonable time from
the sources alone: the reference of Section~\ref{sec:isocout} alone takes 60754
steps on 32256 cells. Two of the checks described above run as regression
tests on every build rather than once. The entropy contract requires a two
hundred step wall bounded rollout to leave the total entropy nonincreasing to
$10^{-10}$. The geometric contracts verify, on each generated case, the domain
area against its analytic value and the count of each boundary face type.

\paragraph{What holds.} Inside a solver whose admissibility does not depend on
the network, learning adds a reproducible gain in distribution and at an unseen
Mach number, and that gain is worth more than the equivalent spending on classical refinement.
What the field is after is the accuracy of a fine mesh at the price of a coarse
one. Read that way, the contribution is worth between $1.12$ and $2.84$
times of classical refinement, for a measured cost of $1.16$ times. The guarantee itself
costs 16\% per step, holds for any weights rather than for trained ones, and
removes a failure mode that neighbouring work reports encountering. On
every case, every geometry and every seed of this study, no negative density or
pressure was produced.

\paragraph*{Acknowledgments and disclosure.} Drafting, editing, and figure
scripting were assisted by an AI system (Claude Opus, Anthropic) operating under
the author's direction. Every number reported here comes from the
version controlled record described in Section~\ref{sec:limites}, in which each
campaign is stored with the thresholds fixed before it was run. Each was checked
against the raw evaluation files. The author reviewed and approves all content.

\bibliographystyle{plainnat}
\bibliography{refs}

\end{document}